\documentclass[
  superscriptaddress,
  twocolumn,
  amsmath,
  amssymb,
  aps,
  prl,
  nofootinbib,
  longbibliography
]{revtex4-2}

\usepackage{graphicx}
\usepackage{dcolumn}
\usepackage{bm}
\usepackage{hyperref}

\usepackage{xspace}
\newcommand{\MeVcc}{MeV/$c^2$\xspace}

\begin{document}

\title{Search for a Higgs Portal Scalar Decaying to Electron-Positron Pairs in the MicroBooNE Detector}

\newcommand{\Bern}{Universit{\"a}t Bern, Bern CH-3012, Switzerland}
\newcommand{\BNL}{Brookhaven National Laboratory (BNL), Upton, NY, 11973, USA}
\newcommand{\UCSB}{University of California, Santa Barbara, CA, 93106, USA}
\newcommand{\Cambridge}{University of Cambridge, Cambridge CB3 0HE, United Kingdom}
\newcommand{\StKates}{St. Catherine University, Saint Paul, MN 55105, USA}
\newcommand{\CIEMAT}{Centro de Investigaciones Energ\'{e}ticas, Medioambientales y Tecnol\'{o}gicas (CIEMAT), Madrid E-28040, Spain}
\newcommand{\Chicago}{University of Chicago, Chicago, IL, 60637, USA}
\newcommand{\Cincinnati}{University of Cincinnati, Cincinnati, OH, 45221, USA}
\newcommand{\CSU}{Colorado State University, Fort Collins, CO, 80523, USA}
\newcommand{\Columbia}{Columbia University, New York, NY, 10027, USA}
\newcommand{\Edinburgh}{University of Edinburgh, Edinburgh EH9 3FD, United Kingdom}
\newcommand{\FNAL}{Fermi National Accelerator Laboratory (FNAL), Batavia, IL 60510, USA}
\newcommand{\Granada}{Universidad de Granada, Granada E-18071, Spain}
\newcommand{\Harvard}{Harvard University, Cambridge, MA 02138, USA}
\newcommand{\IIT}{Illinois Institute of Technology (IIT), Chicago, IL 60616, USA}
\newcommand{\KSU}{Kansas State University (KSU), Manhattan, KS, 66506, USA}
\newcommand{\Lancaster}{Lancaster University, Lancaster LA1 4YW, United Kingdom}
\newcommand{\LANL}{Los Alamos National Laboratory (LANL), Los Alamos, NM, 87545, USA}
\newcommand{\Manchester}{The University of Manchester, Manchester M13 9PL, United Kingdom}
\newcommand{\MIT}{Massachusetts Institute of Technology (MIT), Cambridge, MA, 02139, USA}
\newcommand{\Michigan}{University of Michigan, Ann Arbor, MI, 48109, USA}
\newcommand{\Minnesota}{University of Minnesota, Minneapolis, MN, 55455, USA}
\newcommand{\NMSU}{New Mexico State University (NMSU), Las Cruces, NM, 88003, USA}
\newcommand{\Otterbein}{Otterbein University, Westerville, OH, 43081, USA}
\newcommand{\Oxford}{University of Oxford, Oxford OX1 3RH, United Kingdom}
\newcommand{\Pitt}{University of Pittsburgh, Pittsburgh, PA, 15260, USA}
\newcommand{\Rutgers}{Rutgers University, Piscataway, NJ, 08854, USA}
\newcommand{\SLAC}{SLAC National Accelerator Laboratory, Menlo Park, CA, 94025, USA}
\newcommand{\SDSMT}{South Dakota School of Mines and Technology (SDSMT), Rapid City, SD, 57701, USA}
\newcommand{\Maine}{University of Southern Maine, Portland, ME, 04104, USA}
\newcommand{\Syracuse}{Syracuse University, Syracuse, NY, 13244, USA}
\newcommand{\TelAviv}{Tel Aviv University, Tel Aviv, Israel, 69978}
\newcommand{\Tennessee}{University of Tennessee, Knoxville, TN, 37996, USA}
\newcommand{\UTA}{University of Texas, Arlington, TX, 76019, USA}
\newcommand{\Tufts}{Tufts University, Medford, MA, 02155, USA}
\newcommand{\VTech}{Center for Neutrino Physics, Virginia Tech, Blacksburg, VA, 24061, USA}
\newcommand{\Warwick}{University of Warwick, Coventry CV4 7AL, United Kingdom}
\newcommand{\Yale}{Wright Laboratory, Department of Physics, Yale University, New Haven, CT, 06520, USA}

\affiliation{\Bern}
\affiliation{\BNL}
\affiliation{\UCSB}
\affiliation{\Cambridge}
\affiliation{\StKates}
\affiliation{\CIEMAT}
\affiliation{\Chicago}
\affiliation{\Cincinnati}
\affiliation{\CSU}
\affiliation{\Columbia}
\affiliation{\Edinburgh}
\affiliation{\FNAL}
\affiliation{\Granada}
\affiliation{\Harvard}
\affiliation{\IIT}
\affiliation{\KSU}
\affiliation{\Lancaster}
\affiliation{\LANL}
\affiliation{\Manchester}
\affiliation{\MIT}
\affiliation{\Michigan}
\affiliation{\Minnesota}
\affiliation{\NMSU}
\affiliation{\Otterbein}
\affiliation{\Oxford}
\affiliation{\Pitt}
\affiliation{\Rutgers}
\affiliation{\SLAC}
\affiliation{\SDSMT}
\affiliation{\Maine}
\affiliation{\Syracuse}
\affiliation{\TelAviv}
\affiliation{\Tennessee}
\affiliation{\UTA}
\affiliation{\Tufts}
\affiliation{\VTech}
\affiliation{\Warwick}
\affiliation{\Yale}

\author{P.~Abratenko} \affiliation{\Tufts} 
\author{R.~An} \affiliation{\IIT}
\author{J.~Anthony} \affiliation{\Cambridge}
\author{J.~Asaadi} \affiliation{\UTA}
\author{A.~Ashkenazi} \affiliation{\MIT}\affiliation{\TelAviv}
\author{S.~Balasubramanian}\affiliation{\FNAL}
\author{B.~Baller} \affiliation{\FNAL}
\author{C.~Barnes} \affiliation{\Michigan}
\author{G.~Barr} \affiliation{\Oxford}
\author{V.~Basque} \affiliation{\Manchester}
% ...

\collaboration{The MicroBooNE Collaboration}
\thanks{microboone\_info@fnal.gov}\noaffiliation

\date{\today}

\begin{abstract}
We present a search for the decays of a neutral scalar boson produced by kaons decaying at rest, in the context of the Higgs portal model, using the MicroBooNE detector.
We analyze data triggered in time with the Fermilab NuMI neutrino beam spill, with an exposure of $1.93\times10^{20}$ protons on target.
We look for monoenergetic scalars that come from the direction of the NuMI hadron absorber, at a distance of $100$~m from the detector, and decay to electron-positron pairs.
We observe one candidate event, with a standard model background prediction of $1.9\pm0.8$.
We set an upper limit on the scalar--Higgs mixing angle of $\theta<(3.3-4.6)\times10^{-4}$ at the 95\% confidence level for scalar boson masses in the range $(100-200)$~\MeVcc.
We exclude, at the 95\% confidence level, the remaining model parameters required to explain the central value of a possible excess of $K^0_L\rightarrow\pi^0\nu\bar{\nu}$ decays reported by the KOTO collaboration. We also provide a model-independent limit on a new boson $X$ produced in $K\rightarrow\pi X$ decays and decaying to $e^+e^-$.
\end{abstract}

\maketitle

The Higgs portal model~\cite{higgs-portal-original} is an extension to the standard model in which an electrically neutral real singlet scalar boson ($S$) mixes with the Higgs boson with mixing angle $\theta$.
Through this mixing, $S$ acquires a coupling to standard model fermions proportional to $\sin\theta$ and their Yukawa couplings with the Higgs boson.
For masses between twice the electron mass and twice the muon mass, and assuming that there are no new dark sector particles lighter than half its mass, $S$ will decay to electron-positron pairs with partial width~\cite{hpsbn} 
\begin{equation}
    \Gamma=\theta^2 \frac{m_e^2 m_S}{8\pi v^2}\left(1-\frac{4 m_e^2}{m_S^2}\right)^{\frac{3}{2}},
\end{equation}
where $m_S$ is the scalar boson mass, $m_e$ is the electron mass, and $v$ is the Higgs field vacuum expectation value.
For these masses, $S$ can be produced from a kaon two-body decay in association with a pion, with the dominant production process being a penguin diagram with a top quark running in the loop.
The partial width of the production process is~\cite{hpsbn}
\begin{equation}
    \Gamma\simeq\frac{\theta^2}{16\pi m_K}\left|\frac{3 V_{td}^* V_{ts} m_t^2 m_K^2}{32\pi^2 v^3}\right|^2 \lambda^{1/2}\left(1,\frac{m_S^2}{m_K^2}\frac{m_\pi^2}{m_K^2}\right),
\end{equation}
where $m_K$ is the kaon mass, $m_\pi$ is the pion mass, $m_t$ is the top quark mass, $V_{td}$ and $V_{ts}$ are the elements of the CKM matrix, and $\lambda$ is the K\"allen Lambda function.

In 2019, the KOTO collaboration reported~\cite{koto-anom} the observation of four $K^0_L\rightarrow\pi^0+\textrm{invisible}$ decay candidates, a rate 2 orders of magnitude more frequent than the standard model prediction for $K^0_L\rightarrow\pi^0\nu\bar{\nu}$ decays. One candidate was rejected due to upstream veto activity, but three candidates remained as unexplained.
In a recent publication~\cite{koto-new}, they have reevaluated their background expectation to $1.22\pm0.26$ counts, and the statistical significance of the observed data has reduced to a $p$ value of 0.13. 
The Higgs portal model could explain~\cite{koto-pheno-0,koto-pheno-1,koto-pheno-2} any excess in the KOTO dataset.
The value of $\theta$ for a central value of 1.78 counts ($\theta_\textrm{KCV}$) is ${\approx(4-5)\times10^{-4}}$ over the $m_S$ range of $(100-200)$~\MeVcc.
The E949 collaboration excludes~\cite{e949} $\theta_\textrm{KCV}$ for $m_S<120$~\MeVcc, and the NA62 collaboration excludes it~\cite{na62} for $m_S>160$~\MeVcc.

This Letter presents the first search for beyond the standard model (BSM) electron-positron pair production in a liquid argon time projection chamber using the MicroBooNE detector, and is the second search for BSM physics in MicroBooNE following a search for heavy neutral leptons~\cite{hnl}.
We use the results of this search to exclude, at 95\% confidence level, the remaining Higgs portal model parameter space where $\theta_\textrm{KCV}$ has not been excluded.

The MicroBooNE experiment is primarily designed for neutrino scattering measurements in Fermilab's {\em Booster Neutrino Beam} (BNB)~\cite{bnb,detector}.
The detector sits just below the surface, and comprises an 85~ton liquid-argon time projection chamber (TPC) with active dimensions of 2.6~m along the drift direction (horizontal and perpendicular to the beam axis; $x$ coordinate in the detector reference frame), 2.3~m in the vertical direction ($y$ coordinate), and 10.4~m along the direction parallel to the BNB direction ($z$ coordinate).
Charged particles traversing the argon produce ionization electrons and scintillation light.
Drifted ionization electrons are recorded by three wire planes with 3~mm pitch that are oriented at $60^\circ$ rotations relative to each other.
An array of 32 eight-inch photomultiplier tubes (PMTs) distributed behind the wire planes provides timing information for scintillation signals produced inside the cryostat.
Part way through the detector operations, a cosmic ray tagger (CRT) system~\cite{crt} was installed, with four walls of plastic scintillator panels situated along the top, bottom, and long sides of the cryostat that provide timing coincidence signals for some cosmic rays that enter the TPC.

In addition to being on the BNB beam line, the MicroBooNE detector is also situated at a distance of 680~m and $8^\circ$ off axis from the target of Fermilab's {\em Neutrinos at the Main Injector} (NuMI) neutrino beam~\cite{numibeam}, which we have previously used to measure electron neutrino interactions~\cite{ub-numi-nue}.
\begin{figure}[tp]
    \centering
    \includegraphics[width=8.6 cm]{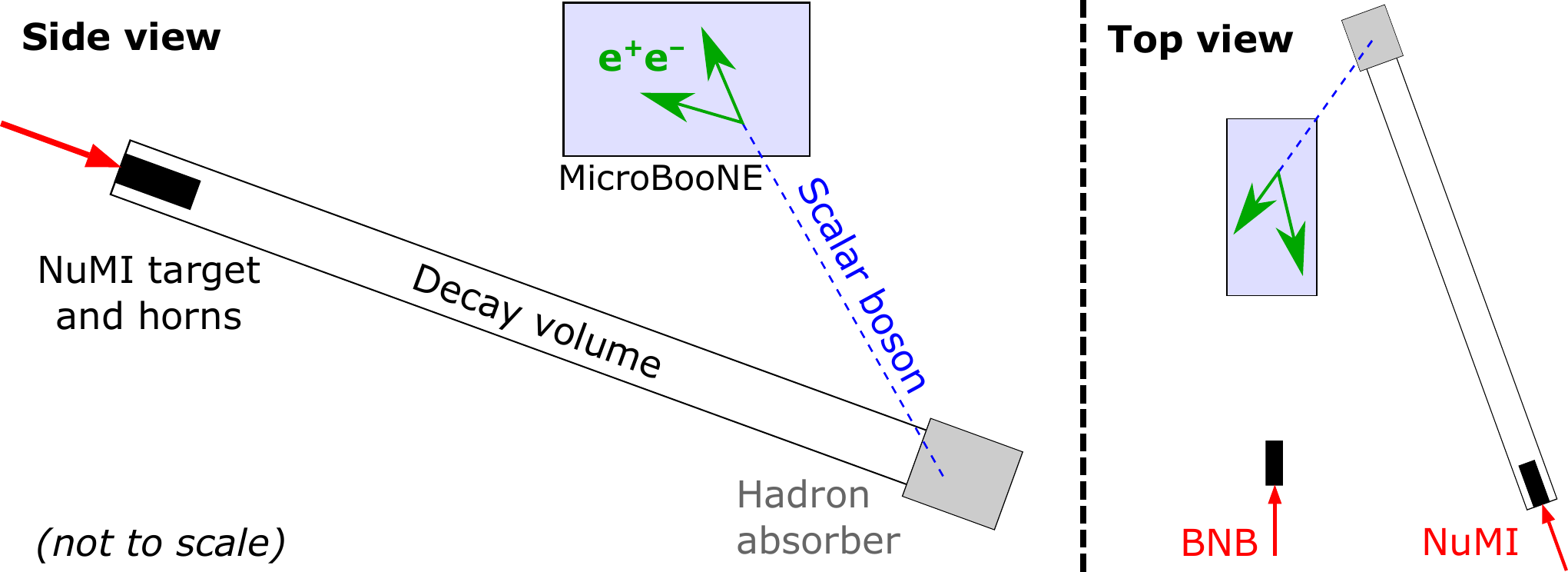}
    \caption{Schematic of MicroBooNE in the NuMI and BNB beam lines, and the signal signature we are searching for.}
    \label{fig:beam}
\end{figure}
A schematic diagram of the detector position within the beam line is presented in Fig.~\ref{fig:beam}.
The Main Injector delivers 120~GeV protons that impact the graphite target, producing secondary hadrons.
A system of electromagnetic horns focuses charged particles either toward or away from the beam axis, depending on the horn polarity.
In forward (reverse) horn current mode, positively (negatively) charged mesons are bent toward the beam axis to produce a beam mostly of neutrinos (antineutrinos) from the meson decays.
A 675~m long helium-filled decay volume is situated downstream of the target and horn system, at the end of which is a 5~m deep hadron absorber.
Any surviving hadrons will be stopped in the absorber, and may produce secondary mesons including $K^+$ which will decay at rest.
The absorber is at a distance of $\approx100$~m from the MicroBooNE detector and at an angle of $\approx125^\circ$ with respect to the BNB direction, such that any particles that enter the detector from the absorber are entering in the opposite direction compared to most neutrino interactions seen by the detector.
We exploit the unique decay signature of scalar bosons produced by $K^+$ decaying at rest (KDAR) in the NuMI hadron absorber to search for evidence of the Higgs portal scalar model.

We analyze $0.92\times10^{20}$~protons on target (POT) of exposure during run~1 of MicroBooNE's operations (during 2015--2016), and $1.01\times10^{20}$~POT of run~3 data (2017--2018).
During the run~1 dataset period, the NuMI beam operated in forward horn current mode, and reverse horn current mode was used during the run~3 period.
The CRT had been fully installed by the run~3  period and we use its information in the analysis of data from that period.
The beam-on data is read out from the detector (an ``event'') when there is a NuMI beam spill timing signal sent by the Fermilab accelerator complex.
An on-line trigger is employed to record only those events that pass optical trigger criteria based on the total integrated charge summed over all PMTs in a 100~ns window.
This trigger requires at least one PMT to produce a signal in time with the beam, and the integrated charge has to be above a configurable photoelectron threshold.

To estimate the cosmic-induced backgrounds, we record a dataset of events produced out of time with both beams that employs the same trigger thresholds, called the beam-off dataset.
In addition, there is an unbiased dataset of out-of-time events for which the trigger is not applied.
This unbiased dataset forms the basis of the simulated data.
The hit pattern of simulated signal decays or background neutrino interactions are overlaid on top of the unbiased data on an event-by-event level, which allows the cosmic contamination of signal or neutrino interactions to be estimated.

To simulate the scalar boson signal, we use the \verb|g4numi| program~\cite{g4numi} which employs a \verb|GEANT4|~\cite{geant4} simulation of the NuMI beam line to produce the position and timing distribution of KDAR in the NuMI hadron absorber.
For the absolute rate, we use the MiniBooNE estimate~\cite{mbkdar} of 0.085 muon neutrinos from KDAR in the NuMI hadron absorber per POT.
The scalars are emitted isotropically from the kaon decay positions, and the scalar's velocity and lab-frame lifetime is used to distribute the scalar decay position, keeping only those that decay within the detector active volume.
The electron-positron pair is simulated isotropically in the rest frame, and boosted by the scalar's momentum.

The \verb|g4numi| program is also used to simulate the flux of neutrinos that intersect the detector, which produce the other component of the background to this search.
We use the same \verb|PPFX|~\cite{g4numi,ppfx} package as used by the MINERvA~\cite{minerva-ppfx} and NOvA~\cite{nova-ppfx} collaborations to correct the central value neutrino flux prediction and provide flux uncertainties.
We use \verb|GENIE|~\cite{genie} to calculate the neutrino interaction cross sections and final state kinematics, in which the models for charged-current quasielastic scattering and scattering on a pair of correlated nucleons have been tuned based on data from T2K~\cite{t2k-genie-tune}. 
Neutrino interactions are simulated both inside the cryostat and outside where secondary products enter the detector. 
For both signal and background simulations, the decay or interaction products are propagated through a \verb|GEANT4| simulation of the detector.
The response of the detector to both light and charge is simulated.

All three types of data (beam on, beam off, and simulated) are processed through the same chain of reconstruction algorithms.
The optical reconstruction uses the PMT waveforms to produce ``flashes'' of coincident PMT hits.
For the TPC information, we apply a two-dimensional deconvolution of the signal waveforms on the wires within each plane~\cite{2deconv,2deconv-2}.
Hits are formed from a Gaussian peak finding algorithm applied to the wire waveform.
The Pandora framework~\cite{pandora} uses particle flow algorithms to cluster the hits of a single plane, and, then, match clusters across planes into three-dimensional reconstructed objects, which Pandora classifies as ``tracks'' or ``showers'' based on a multivariate classifier score.
Pandora also ``slices'' the event into groups of reconstructed objects that it considers to be independent interactions (either cosmogenic or beam induced) and removes well-identified cosmic slices.
For any remaining slices, a flash-matching algorithm is applied to produce a PMT hit hypothesis using the reconstructed objects in the slice.
The algorithm attempts to match the hypothetical PMT hit distribution with the observed flashes in the beam timing window, and calculates a $\chi^2$ value for the best match.
The best matching slice of these remaining slices is labeled as the {\em neutrino slice}.

To preselect decay candidates in the event, we use the neutrino slices.
The slice has to be matched to a PMT flash with a time of $[5.8,16.8]~\mu$s within the $20~\mu$s PMT readout window (where the NuMI prompt neutrino spill produces flashes in the range $[6.1,15.7]~\mu$s), and the flash-matching $\chi^2$ has to be less than 10.
An additional selection is imposed on the run~3 data, requiring that events cannot have a CRT hit in coincidence with the beam timing.
The total number of objects in the slice has to be $\leq5$, and of these, a maximum of 4 can be labeled as tracks.
For all possible pairs of objects in the slice, the minimum distance between the object vertices (for reconstructed tracks, both start or end positions, and for showers, only start positions) is calculated.
If this distance is less than 5~cm, a decay vertex is produced at the midpoint between the object vertices with the minimum separation.
The position of the decay vertex has to be reconstructed within the active volume of the detector.
Slices with more than two objects could conceivably form multiple decay candidates, all of which are preselected and passed through the boosted decision tree (BDT) selection.

We apply two different BDTs to the preselected candidates: one trained against cosmic backgrounds and one trained against neutrino interactions simulated inside the cryostat.
Each BDT is trained separately over the run~1 events and run~3 events, i.e., there are four BDTs in total.
We split the run periods because the use of the CRT in run~3 and the differences between forward and reverse horn current operations can change the topologies and properties of the background distributions that the BDTs are trained against.
We use \verb|xgboost|~\cite{xgboost} to train and apply the BDTs.
We train the BDTs on ten input variables each. 
Nine of the ten input variables are the same for the cosmic-focused and neutrino-focused BDTs.
These are 
(1) the opening angle between the two reconstructed objects; 
(2) the opening angle in the plane transverse to the hadron absorber direction from the detector center; 
(3,4) the two angles between the two objects and the hadron absorber direction; 
(5) the Pandora track or shower score of the larger of the two objects (when ordered by number of hits); 
(6) the number of hits of the larger object; 
(7) the total number of hits contained in other objects in the slice, not including the two objects that form the decay candidate; 
(8) the maximum $y$ coordinate, relative to the decay vertex position, of shower start positions or track start or end positions, for any other objects in the slice; and 
(9) the minimum $z$ coordinate, relative to the decay vertex position, of shower start positions or track start or end positions, for any other objects in the slice.
The last two variables are treated as ``missing'' within \verb|xgboost| if the slice contains only two objects. 
The tenth input variable of the cosmic-focused BDT is the length of the larger object.
The tenth input variable of the neutrino-focused BDT is the number of tracks in the slice.
For all input variables and output BDT score distributions we observe good data-simulation agreement in a control region of data with an early flash time (as the scalar boson signal is delayed by $\approx600$~ns with respect to the prompt neutrino interactions due to time-of-flight differences).

The BDTs are trained on a signal simulation where each decay is of a scalar boson with $m_S$ uniformly chosen between 100 and 200 \MeVcc, in order to reduce the dependence of the BDTs on $m_S$ in the range where $\theta_\textrm{KCV}$ has not been excluded.
The candidates used in the training have to be well reconstructed, with the cosmic contamination of each object below 10\%, and the reconstructed vertex and directions close to the generated values.
The neutrino-focused BDT is trained against 10\% of the simulated statistics of neutrino interactions in the cryostat, with the other 90\% along with all the out-of-cryostat simulated interactions used for the sensitivity and limit calculations.
Each reconstructed object used in the neutrino background training sample is required to have cosmic contamination below 10\%, similar to the signal sample.
The cosmic-focused BDT is trained against beam-off candidates that fail the flash-matching $\chi^2$ requirement.

To select candidate decays, we require the two BDT scores to be above a minimum score. 
We choose the four minimum scores that maximize the sensitivity of the selection to the model parameter $\theta$ for $m_S=100$~MeV/$c^2$, as we expect even better sensitivity at higher masses.
The 95\% confidence level (C.L.) sensitivity and limit are calculated for a single-bin counting experiment with the modified-frequentist CLs method, using the \verb|RooStats| statistical package~\cite{roostats}, including  systematic uncertainties as constrained Gaussian nuisance terms.

We consider several classes of systematic uncertainty.
We include the simulation statistical uncertainty and beam-off data statistical uncertainty as uncertainties for the model prediction.
The flux normalization uncertainty on the signal model is set to $30\%$ as used by MiniBooNE~\cite{mbkdar}.
The uncertainty on the background neutrino cross section modeling is evaluated by reweighting events using tools included with \verb|GENIE|~\cite{genie-rw}.
Interaction physics model parameters (both in \verb|GENIE| and \verb|GEANT4|) are varied multiple times within their 1 standard deviation uncertainties, and a weight is calculated for each simulated event between the central value and the modified model.
The uncertainty on the event count in the selection is calculated from the standard deviation of weighted event counts across the variations. 
A similar procedure is followed to estimate the uncertainty on the background neutrino flux model due to hadron production uncertainties, using \verb|PPFX|~\cite{ppfx}. The flux uncertainty due to the beam line model (including focusing) is negligible compared to the hadron production uncertainty~\cite{ub-numi-nue} and is not included. 

Systematic uncertainties due to the modeling of the detector are evaluated through modified simulations varying parameters of the detector model.
They are estimated to be 70\% for the neutrino background simulation (dominated by the low statistics of simulated neutrino events in the signal region after selection) and 5\% for the signal simulation, taken to be the relative differences of event yields in the signal region between the central value and ten detector model variations summed in quadrature.
The first five detector model variations are (1) uncertainties in the space charge mapping~\cite{spacecharge}, (2) the ion-electron recombination model, (3) a decrease in light yield, (4) an increase in the Rayleigh scattering length, and (5) changing the light attenuation between the anode and cathode sides.
We also modify the simulated TPC wire waveform amplitudes and widths.
The sizes of the modifications are characterized in five dimensions based on hit positions, track angles with respect to the wires, and energy deposited per unit length.
The modification sizes are estimated by comparing orthogonal data samples rich in protons and cosmic muons to the central value simulation.
These five wire modifications (each dimension independently) are then applied to the signal and background simulation and used to extract the event yield variation.
Although the uncertainty on the detector model for the background prediction is large, the final result is statistics-limited, and this uncertainty has minimal impact with respect to repeating the analysis with zero detector uncertainty.
\begin{table}[tb]
    \centering
    \caption{Systematic uncertainties for the signal and background model in the signal region.
    }
    \label{tab:unc}
    \begin{ruledtabular}
    \begin{tabular}{lcc}
    Uncertainty & Background (\%) & Signal (\%) \\
    \hline
    Flux (hadron production) & 26.6 & 30.0\\
    Cross section model & 33.4 & -- \\
    Detector model & 70.0 & 5.0 \\
    Beam-off statistics & 38.0 & -- \\
    Simulation statistics & 28.2 & $< 2.0$ \\
    \end{tabular}
    \end{ruledtabular}
\end{table}
The uncertainties in the signal region after the optimal BDT selection are given in Table~\ref{tab:unc}.

\begin{table}[tb]
    \centering
    \caption{Estimated signal selection efficiency (eff.) for a scalar boson decay inside the TPC, and event yield [unweighted (unwt.) and beam-on exposure-weighted (exp. wt.), with the expected signal for $\theta_\textrm{KCV}$].}
    \begin{ruledtabular}
    \begin{tabular}{cccc}
    & & \multicolumn{2}{c}{Event count}\\\cline{3-4}
    Category & Eff. (\%) & Unwt. & Exp. Wt. \\
    \hline
    Beam-off dataset    & & 10   & $1.1\pm0.4$ \\
    Neutrino simulation & & 16   & $0.8\pm0.7$ \\
    Signal (120~\MeVcc) & $14.0\pm0.8$ & 7268 & $4.9\pm1.5$ \\
    Signal (160~\MeVcc) & $14.9\pm0.9$ & 7654 & $12.2\pm3.6$ 
    \end{tabular}
    \end{ruledtabular}
    \label{tab:eventyields}
\end{table}
After applying the BDT selection, the number of events expected for each background contribution and for several signal definitions are shown in Table~\ref{tab:eventyields}. The table also presents the estimated signal selection efficiency.
The total expected background-only prediction is $1.9\pm0.8$ candidate events.

In the beam-on dataset, we observe two candidates in the signal region.
We reject one candidate because its flash time of $5.84~\mu$s lies in the window between the start of the selection time ($5.8~\mu$s) and the start of the neutrino interactions ($6.1~\mu$s), making it an obvious cosmic background interaction. This post-selection cut only affects the cosmic background, reducing the cosmic acceptance by 2.7\%, with negligible effect on the sensitivity.
When we manually inspect the TPC readout of the other candidate event, the two objects have the characteristics of a proton and a photon, and so, it is likely to be a neutrino-induced background.

\begin{figure}[tb]
    \centering
    \includegraphics[width=8.6 cm]{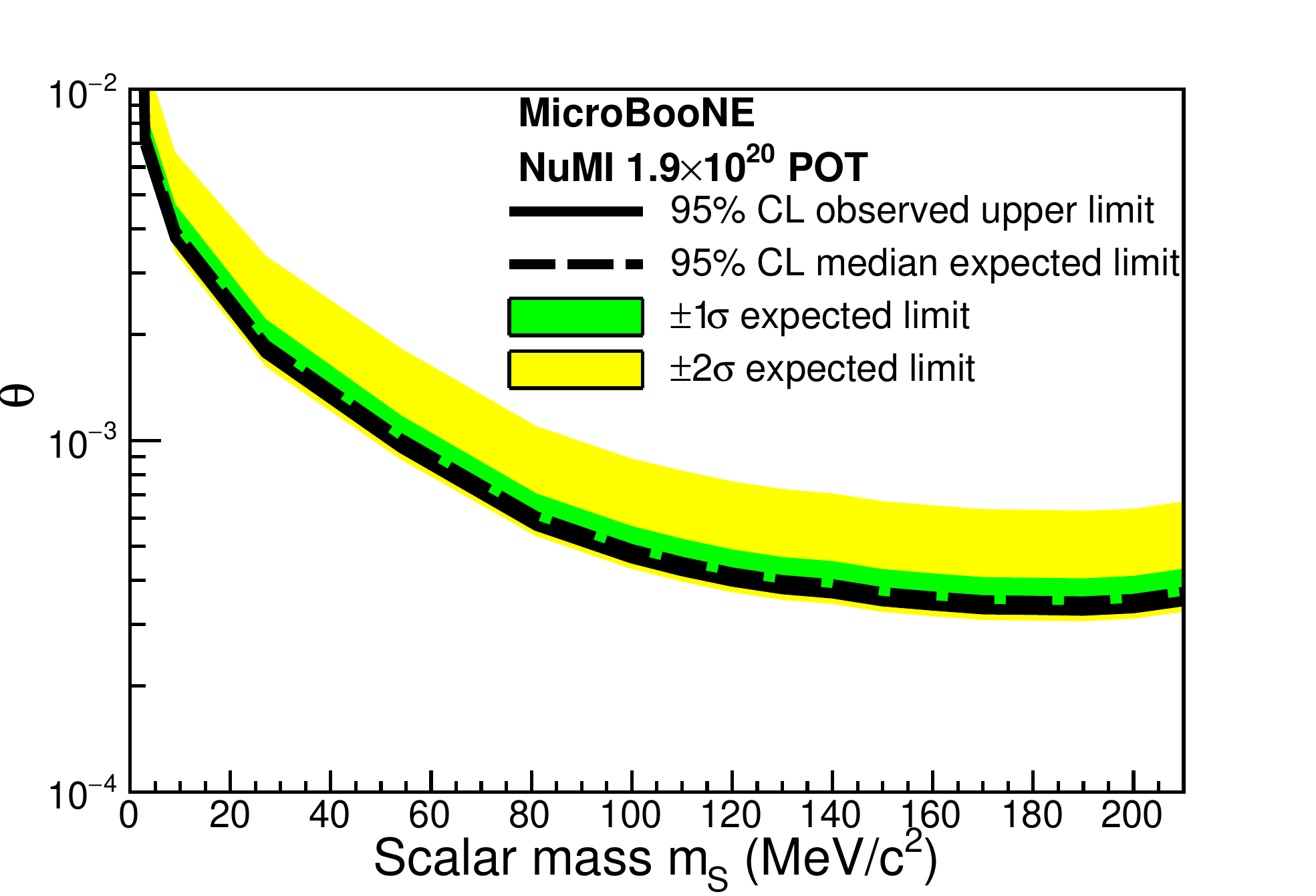}
    \caption{The 95\% confidence level sensitivity and observed limit of this search to the Higgs portal model parameter $\theta$.}
    \label{fig:sens}
\end{figure}
With one observed event we set the 95\% C.L. upper limit on the Higgs portal model presented in Fig.~\ref{fig:sens}.
\begin{table}[tb]
    \centering
    \caption{Observed (obs.) and expected [median (med.), $\pm1,2$ standard deviation] 95\% C.L. upper limits on the Higgs portal parameter $\theta$ for several scalar boson masses.}
    \begin{ruledtabular}
    \begin{tabular}{ccccccc}
    $m_S$ & Obs. & \multicolumn{5}{c}{Expected limits ($10^{-4}$)} \\\cline{3-7}
    (\MeVcc)  & limit ($10^{-4}$) & $-2$ s.d. & $-1$ s.d. & med. & $+1$ s.d. & $+2$ s.d. \\
    \hline
    120 & 4.0 & 3.7 & 3.9 & 4.2 & 4.9 & 7.6 \\
    140 & 3.7 & 3.4 & 3.6 & 3.9 & 4.5 & 7.1 \\
    160 & 3.4 & 3.2 & 3.3 & 3.6 & 4.2 & 6.5 \\
    \end{tabular}
    \end{ruledtabular}
    \label{tab:limits}
\end{table}
The observed and expected limits for several scalar boson masses are enumerated in Table~\ref{tab:limits} and, for a wider range of masses, in the Supplemental Material~\cite{suppl}. 
\begin{figure}[tb]
    \centering
    \includegraphics[width=8.6 cm]{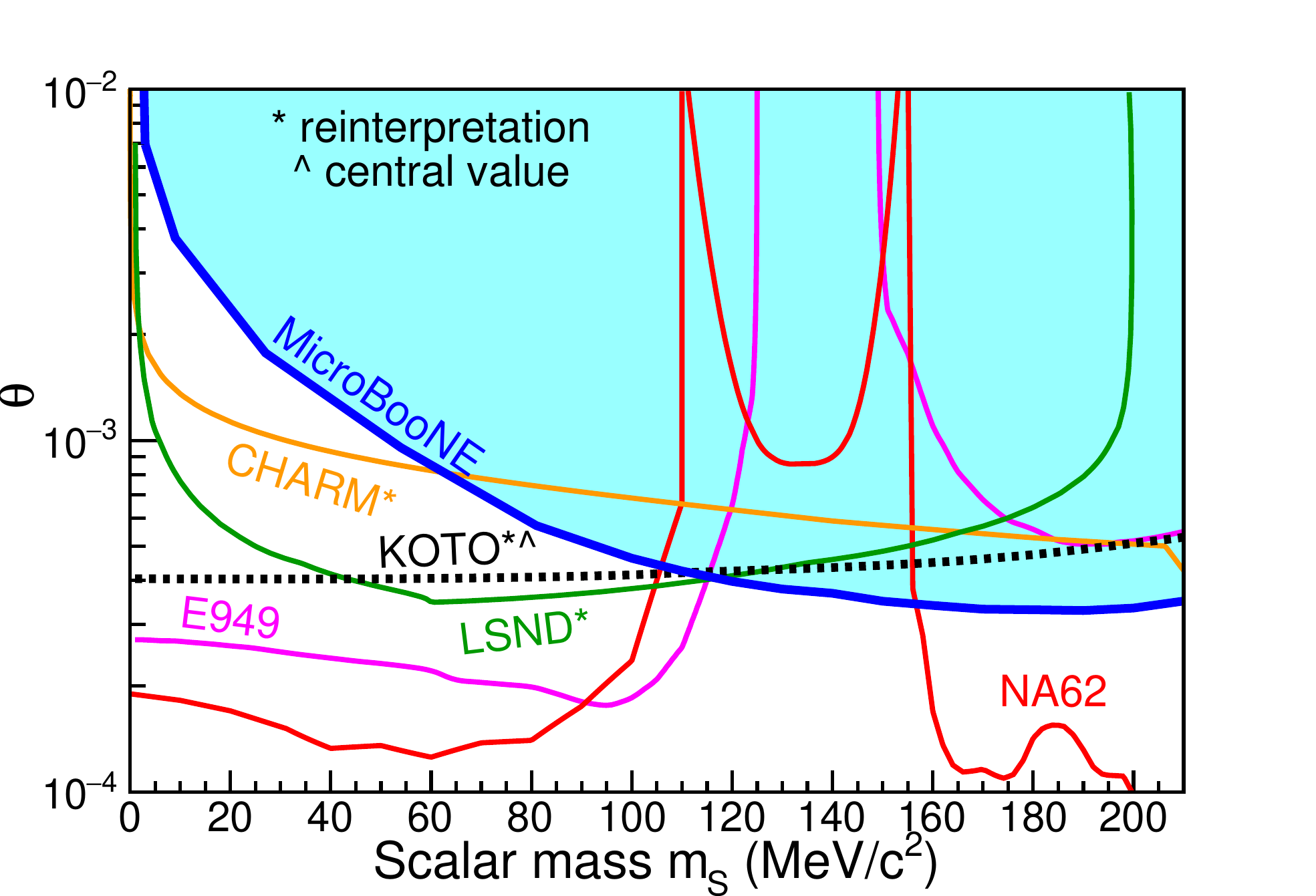}
    \caption{The MicroBooNE 95\% C.L. upper limit (shaded) in the context of the model parameter $\theta_\textrm{KCV}$ required to explain the central value of 1.78 counts in KOTO (dotted line), and exclusion contours from other experiments (solid lines; regions above the lines are excluded). The KOTO central value is adapted from Ref.~\cite{koto-pheno-1} and scaled by $\sqrt{(3-1.22)/3}$ to reflect updated background estimates from the KOTO collaboration~\cite{koto-new}. The limits for E949~\cite{e949} and NA62~\cite{na62} are published by the collaborations, whereas the CHARM~\cite{koto-pheno-1} and LSND~\cite{lsnd} limits are reinterpretations of other searches.}
    \label{fig:limext}
\end{figure}
The upper limit is compared with $\theta_\textrm{KCV}$, along with other experimental limits, in Fig.~\ref{fig:limext}.
\begin{figure}[tb]
    \centering
    \includegraphics[width=8.6 cm]{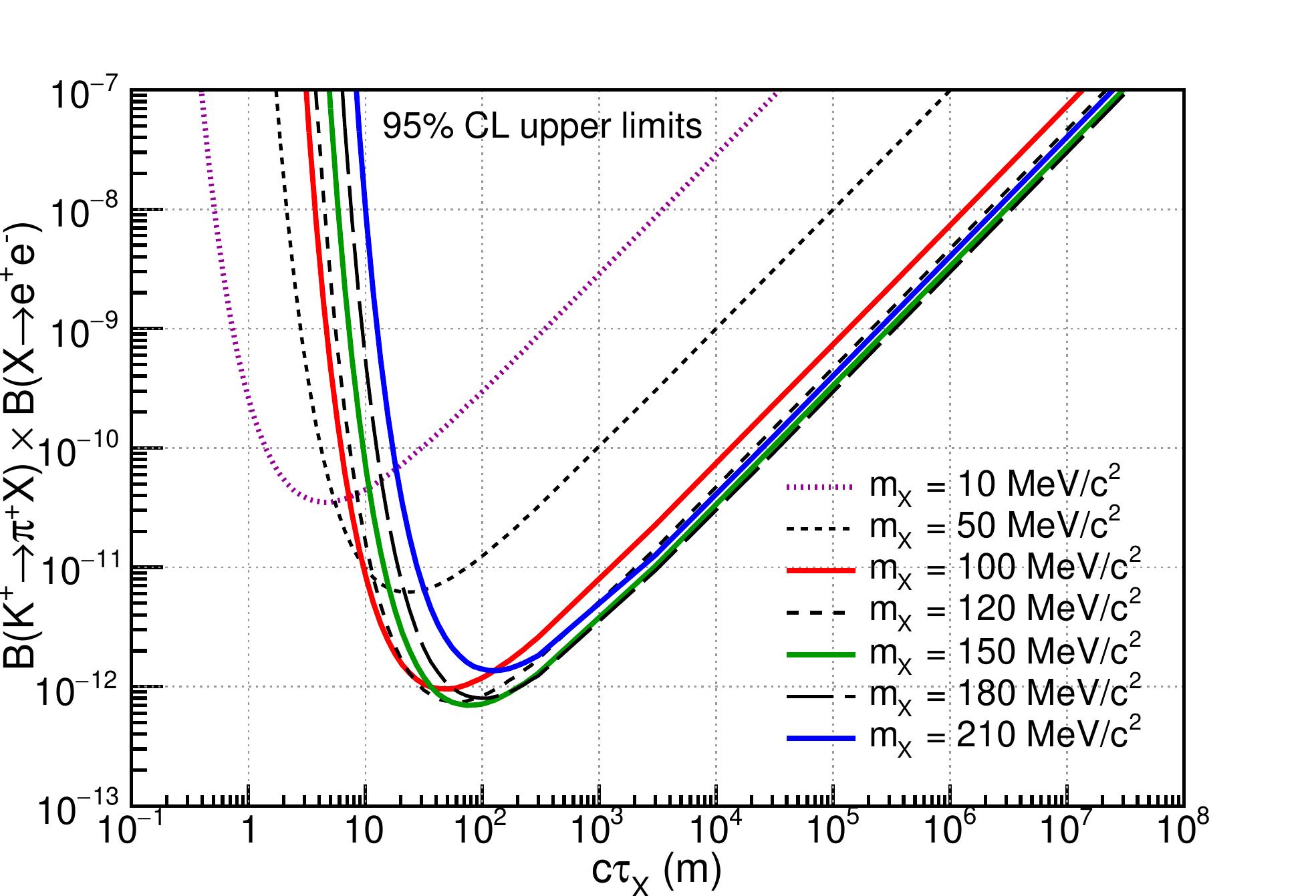}
    \caption{Model-independent upper limits on the product of the production and decay branching ratios of a new boson $X$ as a function of the $X$ boson lifetime $\tau_X$ and mass $m_X$.}
    \label{fig:modindep}
\end{figure}
In Fig.\ref{fig:modindep}, we present our result as a model-independent limit on a new boson $X$ produced in $K^+\rightarrow\pi^+X$ decays, and decaying to $e^+e^-$ pairs, for $X$ masses in the range $[100,210]$~\MeVcc.

The limit presented in this publication rules out the remaining Higgs portal model parameter space required to explain the central value of a mild excess in KOTO at the 95\% confidence level.
Our limit is the most constraining for ${m_S\approx(120-160)}$~\MeVcc and is directly derived from our own experimental data.
The previous most stringent constraints in this range were reinterpretations of decades-old CHARM~\cite{charm,koto-pheno-1} and LSND~\cite{lsnd} measurements, performed recently by independent authors without access to the raw experimental data.
We have $\approx2\times10^{21}$~POT of as-yet unprocessed NuMI data along with $\approx1\times10^{21}$~POT of currently blinded BNB data that we will analyze in the future and expect improved sensitivities~\cite{hpsbn}.

This document was prepared by the MicroBooNE collaboration using the resources of the Fermi National Accelerator Laboratory (Fermilab), a U.S. Department of Energy, Office of Science, HEP User Facility.
Fermilab is managed by Fermi Research Alliance, LLC (FRA), acting under Contract No. DE-AC02-07CH11359.
MicroBooNE is supported by the following: the U.S. Department of Energy, Office of Science, Offices of High Energy Physics and Nuclear Physics; the U.S. National Science Foundation; the Swiss National Science Foundation; the Science and Technology Facilities Council (STFC), part of the United Kingdom Research and Innovation;  and The Royal Society (United Kingdom).
Additional support for the laser calibration system and cosmic ray tagger was provided by the Albert Einstein Center for Fundamental Physics, Bern, Switzerland.
Participation of individual researchers in this project is supported by funds from the European Union’s Horizon 2020 Research and Innovation Programme under the Marie Sk\l{}odowska-Curie Grant Agreement No. 752309.
We thank Brian Batell, Joshua Berger, and Ahmed Ismail for outlining the search strategy. 

\bibliography{biblio}

\end{document}